\documentstyle[twoside,fleqn,espcrc2,epsf]{article}

\newcommand{\be}{\begin{equation}}
\newcommand{\ee}{\end{equation}}

\newcommand{\AmS}{{\protect\the\textfont2
  A\kern-.1667em\lower.5ex\hbox{M}\kern-.125emS}}

\title{Heavy-Light Meson Semileptonic Decays with Staggered Light Quarks}

\author{J.Shigemitsu\address{Physics Department, The Ohio State
        University, Columbus, OH 43210, USA.},
        C.T.H.Davies$^{\rm b}$, A.Gray\address{Department of Physics \&
               Astronomy, University of Glasgow, Glasgow, G12 8QQ, UK.},
        E.Gulez$^{\rm a}$,
       G.P.Lepage\address{Laboratory of Elementary Particle Physics,
        Cornell University, 
                 Ithaca, NY 14853, USA.}, M.Wingate$^{\rm a}$
              }

\begin{document}

\begin{abstract}
  We report on exploratory studies of heavy-light meson semileptonic 
  decays using AsqTad light quarks, NRQCD heavy quarks and Symanzik 
  improved glue on coarse quenched lattices. Oscillatory contributions 
  to three-point correlators coming from the staggered light quarks 
  are found to be handled well by Bayesian fitting methods. B meson 
  decays to both the Goldstone pion and to one of the point-split
  non-Goldstone pions are investigated.  One-loop perturbative 
  matching of NRQCD/AsqTad heavy-light currents is incorporated.
\vspace{1pc}
\end{abstract}

\maketitle

\section{Introduction}
A major systematic error in many studies of heavy-light meson 
decays has been the uncertainty due to chiral extrapolations in the light 
valence and light dynamical quark masses.   These chiral extrapolation 
errors can be reduced significantly by  working 
 with light quark actions that have good 
chiral properties and this has motivated us to
 initiate a program to study heavy-light physics
 using improved staggered (or equivalently improved naive) light quarks 
coupled with NRQCD heavy quarks \cite{hlstagg}.
  Our first results on $B$ and $D$ 
meson decay constants on the MILC dynamical configurations were presented
by M.Wingate at this conference \cite{matt}.  Here we discuss 
initial semileptonic decay results.  

Our semileptonic studies 
are at a much more preliminary stage compared to the decay constant 
calculations. We report on quenched simulations with light quark mass 
around $m_{strange}$ and the heavy quark mass around the $b$-quark 
mass.  
We employ the Asqtad light quark action, the $O(1/M)$ and 
$O(a^2)$ improved NRQCD heavy quark action and the tree-level 
Symanzik improved glue action.  We work with 200 
$12^3 \times 32$ quenched configurations with $a^{-1} \approx 1.0$ GeV.

\section{ Fitting Oscillatory Three-point Correlators}
Two- and three-point correlators involving naive or staggered 
propagators have contributions that oscillate with time and this makes it
 more challenging to carry out fits and extract groundstate contributions.
Experience gained recently with NRQCD/staggered 
 heavy-light two-point correlators have demonstrated the usefulness of 
Bayesian fitting methods \cite{hlstagg,bayes}. 
 Hence, much of the effort in this exploratory 
study of semileptonic decays has focused on testing Bayesian fits 
to oscillatory three-point correlators. The basic correlator of 
interest is,
\begin{eqnarray}
& & C^{(3)}(\vec{p}_\pi, \vec{p}_B, t, T_B)   =   \nonumber \\  
 & & \sum_{\vec{z}} \sum_{\vec{y}} \langle \; \Phi_\pi(0) 
J^\mu(\vec{z},t) \Phi^\dagger_B(\vec{y},T_B) \; \rangle \nonumber \\
& & \qquad \qquad \times 
\, e^{i\vec{p}_B\cdot \vec{y}} \, e^{i(\vec{p}_\pi - \vec{p}_B)\cdot \vec{z}}
\end{eqnarray}
In the present simulations the $B$ meson decays at rest ($\vec{p}_B = 0$) 
and calculations have been performed with $T_B=12$ and $T_B=16$. The 
location of the current insertion $t$ was varied between $1$ and $T_B$. 
The three-point correlator is fit to the form, 
\newpage
\begin{eqnarray}
&& C^{(3)}(\vec{p}_\pi, \vec{p}_B, t, T_B)   \rightarrow 
  \nonumber \\  
 & &  \sum_{k=0}^{N_\pi-1} \sum_{j=0}^{N_B-1}
 (-1)^{k*(t-1)} \, (-1)^{j*(T_B-t)} 
 \qquad \qquad \qquad \nonumber \\
&& \qquad \quad \times  A_{j,k}
\,  e^{-E_\pi^{(k)} (t-1)} \, e^{ -E_B^{(j)} (T_B-t)}
\end{eqnarray}
In this ansatz every second exponential 
 comes with an oscillatory amplitude 
in both the pion and the 
$B$ channels,
  We were able to get 
good fits with $N_\pi = 1$ or $2$ and $N_B = 3$ to $8$.
 Furthermore we found it useful to do simultaneous fits 
to $C^{(3)}$ and to the $B$ meson two-point correlator. Fits were 
done to all data between $t=2$ and $t = T_B - 1$, i.e. to all but 
the first and the last timeslices,  those timeslices on which the 
pion and the $B$ meson  respectively were created.
Figs. 1 \& 2 
give examples of fits to matrix elements of the temporal 
($V_0$) component of the heavy-light 
vector current.  The plots show,
$ C^{(3)}(\vec{p}_\pi,t,T_B) 
\;e^{E^{(0)}_\pi (t-1) } \; e^{E^{(0)}_B(T_B-t)}$. 
By factoring out the groundstate contributions, the 
 opposite parity oscillatory terms and other 
excited state contributions show up more clearly.
Despite the presence of  these excitations, the Bayesian fitting 
methods allow us to extract the groundstate amplitude of interest,
$A_{00}$ in eq.(2). 

 The two figures are for the cases when the pion 
has momentum $(0,1,1)$ in units of $2 \pi/(aL)$.  Good $\chi^2/dof$ 
were found with $N_\pi = 1$ and $N_B = 6$.  The difference between
 the $T_B=12$ and $T_B=16$ results is typical for all matrix elements.
With $T_B=12$, statistical errors are reduced and more accurate fits 
can be made. On the other hand, with $T_B=16$ a hint of a plateau can 
be observed (which will likely disappear with higher statistics 
and a more expanded vertical scale). Since with 
multi-exponential Bayesian fits
 the presence of a plateau
is not required, one is better off with the more accurate 
$T_B=12$ data.  Nevertheless, working with two different $T_B$ values
 provides a good consistency check, which we exploit in the next section.

\begin{figure}
\epsfxsize=7.0cm
\centerline{\epsfbox{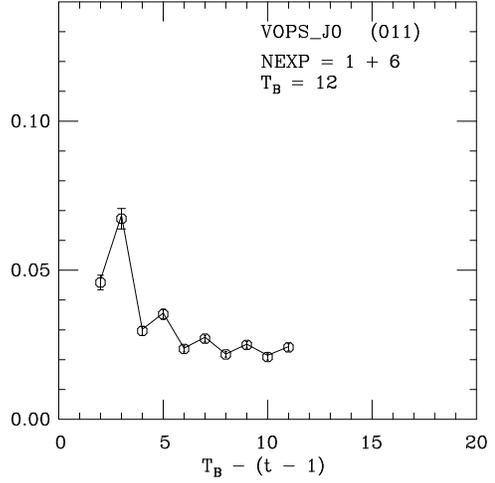}}
\caption{Fit results for the groundstate amplitude $A_{00}$ from 
the $\langle V_0 \rangle$ three-point correlator with $T_B=12$. 
The pion momentum 
is $(0,1,1)2\pi/(aL)$. $N_{exp}$ gives $N_\pi + N_B$.
 }
\end{figure}

\begin{figure}
\epsfxsize=7.0cm
\centerline{\epsfbox{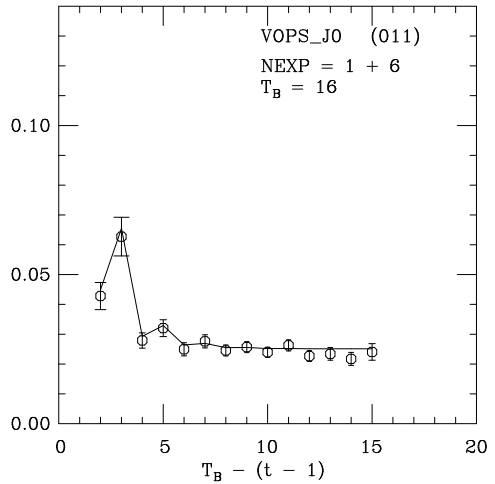}}
\caption{
Same as Fig.1 but for $T_B=16$.
 }
\end{figure}

\section{Results for Form Factors }
Given the amplitudes $A_{00}$ from fits to three-point 
correlators one can determine various form factors. In these 
 preliminary investigations we find it most convenient 
 to work with the form factors $f_{\|}$ and $f_{\perp}$ defined as,
\begin{eqnarray}
f_{\|} & = & \frac{ A_{00}(V_0)}
{\sqrt{\zeta^{(0)}_\pi \zeta^{(0)}_B}}
\, \sqrt{2 E_\pi} \, Z_{V_0}   \\
f_{\perp} & = & \frac{ A_{00}(V_k)}
{\sqrt{\zeta^{(0)}_\pi \zeta^{(0)}_B}
\,p_{\pi,k} }
\, \sqrt{2 E_\pi} \, Z_{V_k}  
\end{eqnarray}
$\zeta^{(0)}_\pi$ and $ \zeta^{(0)}_B$ denote the groundstate amplitudes 
of the pion and $B$ correlators respectively. The $Z_{V_\mu}$ are the matching 
coefficients for the heavy-light currents.  We use one-loop perturbative 
results for them.

In Fig.3 we show results for $f_{\|}$ and $f_{\perp}$.  Our main results 
are the octagons and diamonds, which correspond to $T_B=12$ and $T_B=16$ 
results respectively, for decays into local Goldstone 
pions.  One sees consistency between the two $T_B$ results, with, 
as expected, smaller statistical (and fitting) errors when $T_B=12$.

We have also considered $B$ decays to one of the point-split 
non-Goldstone pions, namely the one-link pion.  
The masses of the one-link and local pions differ as
$M_\pi(one-link)/M_\pi(local) = 1.13$.  The ``bursts'' in Fig.3 show 
form factors for the decay to the one-link pion. One sees that 
at low momentum agreement is reasonable with the local pion results. Things 
start to differ at higher pion momentum, however the data has started 
to deteriorate at those momenta and one also expects lattice artifacts 
to be worse there.  Things should improve considerably as one goes to 
 lattices finer than the $a^{-1} \approx 1$GeV lattices of the present study.
One can conclude from Fig.3 that taste-breaking lattice artifact errors 
are smaller in form factors than in the corresponding pion masses, 
particularly at low pion momenta.

The above form factor results come from the lowest order (in $1/M$) 
heavy-light currents.  We have also analysed matrix elements of all the 
dimension 4 current corrections that come in at the next order (two 
additional currents for $V_0$ and four more for $V_k$).  We find that 
their contributions are small for $B$  mesons giving at most a 
$\sim 2 \; - \; 3$\% correction.

\begin{figure}
\epsfxsize=7.0cm
\centerline{\epsfbox{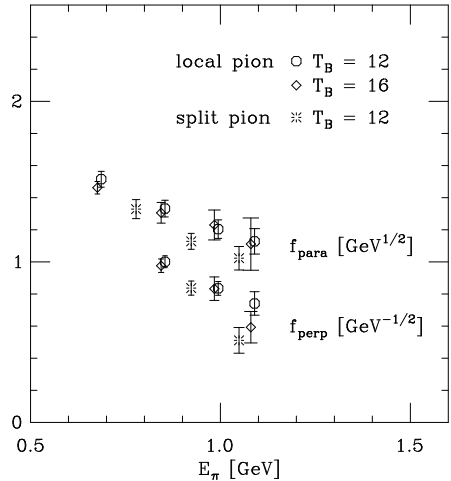}}
\caption{
Semileptonic decay form factors $f_{\|}$ and $f_{\perp}$ versus
the pion energy $E_\pi$ in the $B$ rest frame. 
 }
\end{figure}

\section{Summary }
Exploratory quenched studies of heavy-light meson semileptonic decays 
with Asqtad light quarks look promising.  Simulations are underway 
to calculate form factors on the MILC dynamical configurations.

\vspace{.1in}
\noindent
Acknowledgements : This work was supported by the DOE, 
 PPARC and NSF. Simulations were carried out 
 at the Ohio Supercomputer Center.


\begin{thebibliography}{99}

\bibitem{hlstagg}
M. Wingate {\em et al.}; Phys. Rev. {\bf D67}, 054505 (2003). 

\bibitem{matt}
M. Wingate {\em et al.}; these proceedings and in preparation.

\bibitem{bayes}
G.P. Lepage {\em et al.}; 
 Nucl. Phys. {\bf B}(Proc. \ Suppl.){\bf 106}, 12 (2002).



\end{thebibliography}
\end{document}